\documentclass[twoside,a4paper,11pt]{torus2012}
\usepackage{graphicx}
\usepackage{hyperref}
\usepackage{movie15}
\usepackage{verbatim,amssymb}
\begin{document}
\pagenumbering{arabic}
\pagestyle{myheadings}
\thispagestyle{empty}
{\flushleft\includegraphics[width=\textwidth,bb=58 650 590 680]{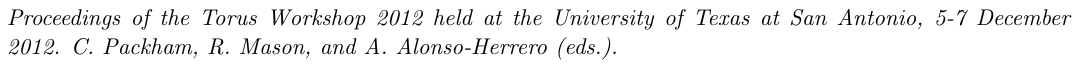}}
\vspace*{0.2cm}
\begin{flushleft}
{\bf {\LARGE
%
Probing The Central Engines Of Low-Luminosity Active Galactic Nuclei
%
}\\
\vspace*{1cm}
%
Rodrigo Nemmen$^{1}$
%
}\\
\vspace*{0.5cm}
%
$^{1}$
NASA Goddard Space Flight Center, Greenbelt, MD 20771, USA. 
%
\end{flushleft}
%
\markboth{
Probing Low-luminosity AGNs
}{ 
%
Nemmen
%
}
\thispagestyle{empty}
\vspace*{0.4cm}
\begin{minipage}[l]{0.09\textwidth}
\ 
\end{minipage}
\begin{minipage}[r]{0.9\textwidth}
\vspace{1cm}
\section*{Abstract}{\small
%
I give a brief review of the observational properties of low-luminosity AGNs (LLAGNs).  I outline some unresolved issues in the study of LLAGNs, emphasizing the uncertainties in the role of the truncated thin accretion disk, the dusty obscuring torus and the origin of high-energy radiation (X-rays and $\gamma$-rays). I discuss key future directions for progress, focusing on the broadband multiwavelength observations that will help us address these issues and the importance of high-resolution infrared observations. 
%
\normalsize}
\end{minipage}
%
%
%
\section{Introduction \label{intro}}

Most active galactic nuclei (AGN) in the present-day universe are characterized by low-luminosities and are thus called low-luminosity AGNs (LLAGNs; \cite{ho97,nagar05,ho08}). 
The bulk of the LLAGN population ($\approx 2/3$; \cite{ho08,ho09}) resides in low-ionization nuclear emission-line regions (LINERs; \cite{heckman80,ho97}). LLAGNs are extremely sub-Eddington systems which are many orders of magnitude less luminous than quasars, with average bolometric luminosities $\left \langle L_{\rm bol} \right \rangle \sim 10^{40}-10^{41} \ {\rm erg \ s}^{-1}$ and an average Eddington ratio of $L_{\rm bol}/L_{\rm Edd} \sim 10^{-5}$ \cite{ho09} where $L_{\rm Edd}$ is the Eddington luminosity. Therefore, it is fair to say that most supermassive black holes in nearby galaxies are ``taking a nap''.

The observational properties of LLAGNs are quite different from those of more luminous AGNs. Regarding the broadband spectral energy distributions (SEDs), LLAGNs seem not to have the thermal continuum prominence in the ultraviolet (UV) -- the ``big blue bump'' -- which is one of the signatures of the presence of an optically thick, geometrically thin accretion disk \cite{ho99, erac10, younes12,nemmen12}. 
Furthermore, with the typical fuel supply of hot diffuse gas (via Bondi accretion) and cold dense gas (via stellar mass loss) available in nearby galaxies, LLAGNs would be expected to produce much higher luminosities than observed on the assumption of standard thin disks with a $10\%$ radiative efficiency \cite{ho09}.
Taken together, this set of observational properties favors the scenario in which the accretion flow in LLAGNs is advection-dominated or radiatively inefficient.

Advection-dominated accretion flows (ADAFs\footnote{Here we use the term ADAF as a synonymous to radiatively inefficient accretion flows (RIAFs).}; 
\cite{nar98,yuan07,nar08}) are very hot, geometrically thick, optically thin flows characterized by low radiative efficiencies ($L \ll 0.1 \dot{M} c^2$) and occur at low accretion rates ($\dot{M} \lesssim 0.01 \dot{M}_{\rm Edd}$). SMBHs are thought to spend most of their lives in the ADAF state \cite{hopkins06,xu10}, the best studied individual case being Sgr A* which is in fact the nearest LLAGN (e.g., \cite{yuan03,yuan07,broder11,scherba12}). On this review, I will concentrate on extragalactic LLAGNs.

In many LLAGNs, another component in the accretion flow besides the ADAF is required in order to account for different observations including a ``red bump'' in the SEDs (e.g., \cite{ho99,nemmen06,yu11,nemmen12}: the emission from a thin accretion disk whose inner radius is truncated at the outer radius of the ADAF. The accretion flow may begin as a standard thin disk depending on the boundary conditions at the Bondi radius but somehow at a certain transition radius it can gradually switch from a cold disk to the hot ADAF mode (e.g., \cite{yuan04,nar08}), similarly to what is required to explain the transition between the different spectral states in black hole binary systems \cite{done07}. The physics of this transition is not very clear (e.g., \cite{nar08}).

ADAFs seem to be quite efficient at producing outflows and jets \cite{nemmen07, sasha11} such that we would expect an association between the LLAGN phenomenon and jets. Indeed, this is supported by observations showing that (i) LLAGNs are usually radio-loud \cite{ho08}, (ii) jets from LLAGNs carry significant amounts of kinetic power \cite{merloni07, heinz07}, and (iii) core LLAGN radio emission is well explained by jet models \cite{yu11,nemmen12} (but see \cite{liu13}).

\begin{figure}
\center
\includegraphics[scale=0.3]{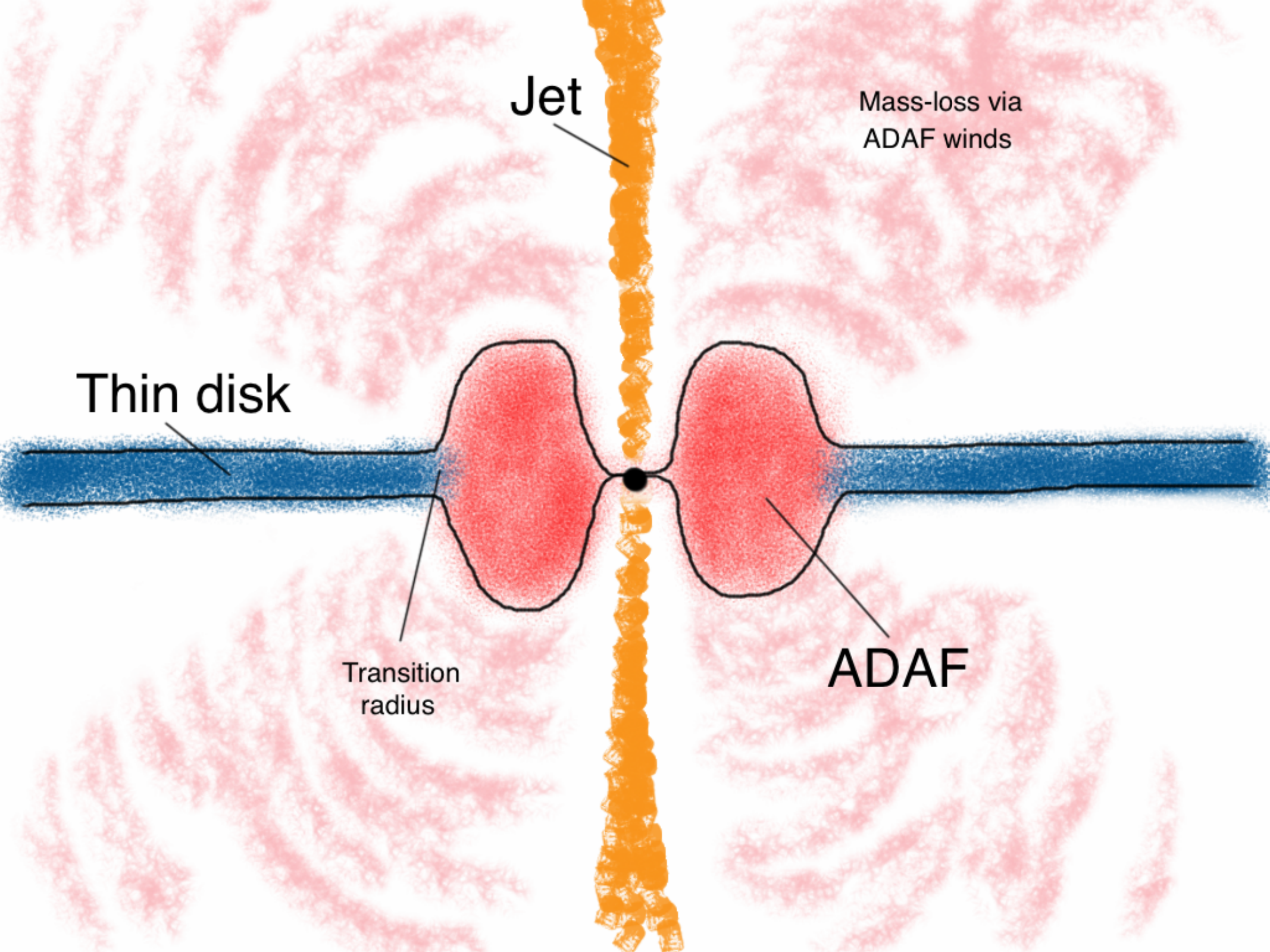}
\caption{Cartoon illustrating the model for the central engines of LLAGNs. It consists of three components: an inner ADAF, an outer truncated thin disk and a relativistic jet.}
\label{cartoon} 
\end{figure}

The ADAF model -- together with an outer truncated thin accretion disk component and a relativistic jet  as illustrated in Figure \ref{cartoon} -- is the theoretical framework that naturally accounts for the set of observational properties of LLAGNs \cite{yuan07,nar08,ho08,yu11,nemmen12}. Hence, we favor this physical scenario in this review. 

\section{Putting ADAFs to the test: Broadband spectra of LLAGNs}

The hot plasma in the ADAF produces a complex broadband spectrum from radio to $\gamma$-rays which includes a combination of photons produced in Synchrotron, bremsstrahlung, neutral pion decay as well as the inverse Compton scattering of synchrotron photons (see Figure 5 in \cite{nar98}). The jet and truncated thin disk contribute to this complex mix of photons with  synchrotron light and the thermal emission, respectively.  
How is this rich variety of radiative processes constrained by the available broadband spectral energy distributions (SEDs) of LLAGNs?

\begin{figure}
\centerline{
\includegraphics[scale=0.5]{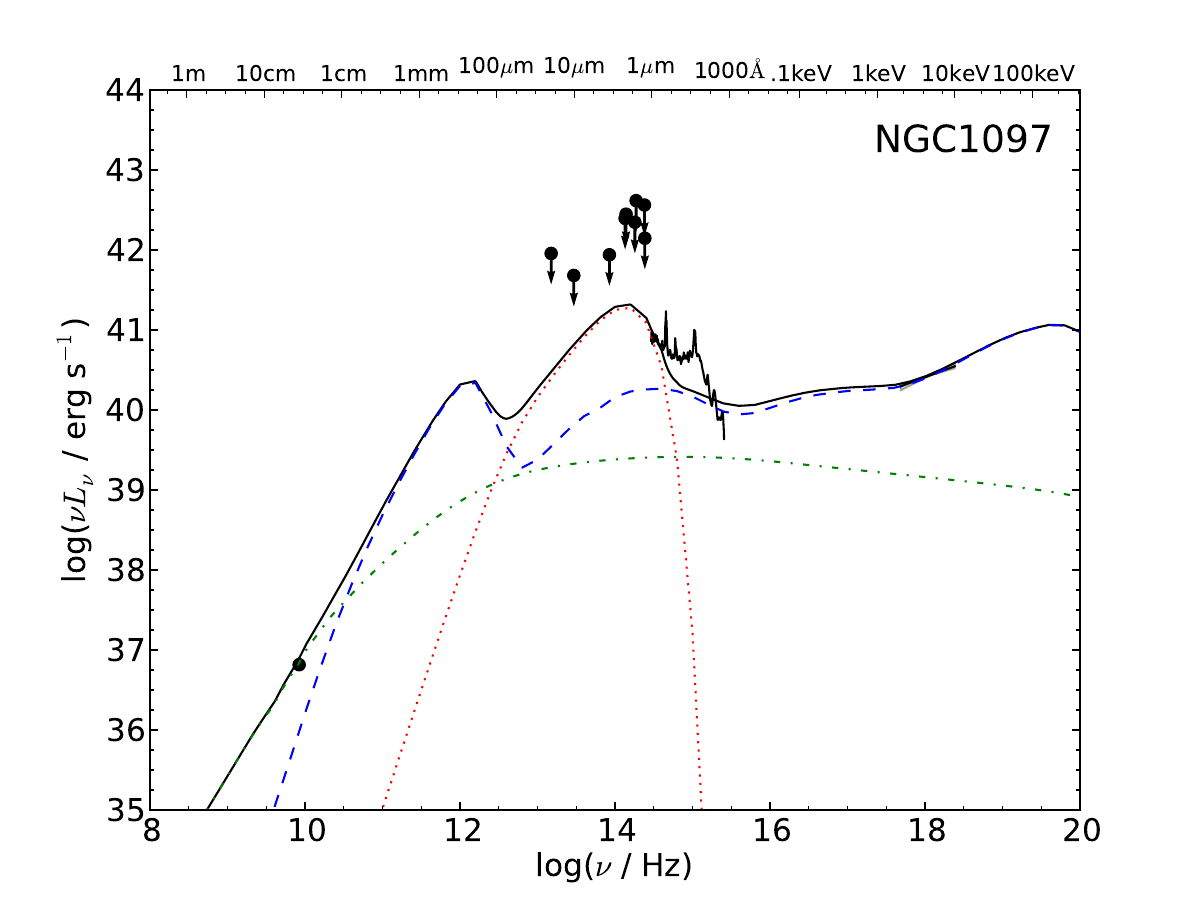}
\hskip -0.3truein
\includegraphics[scale=0.5]{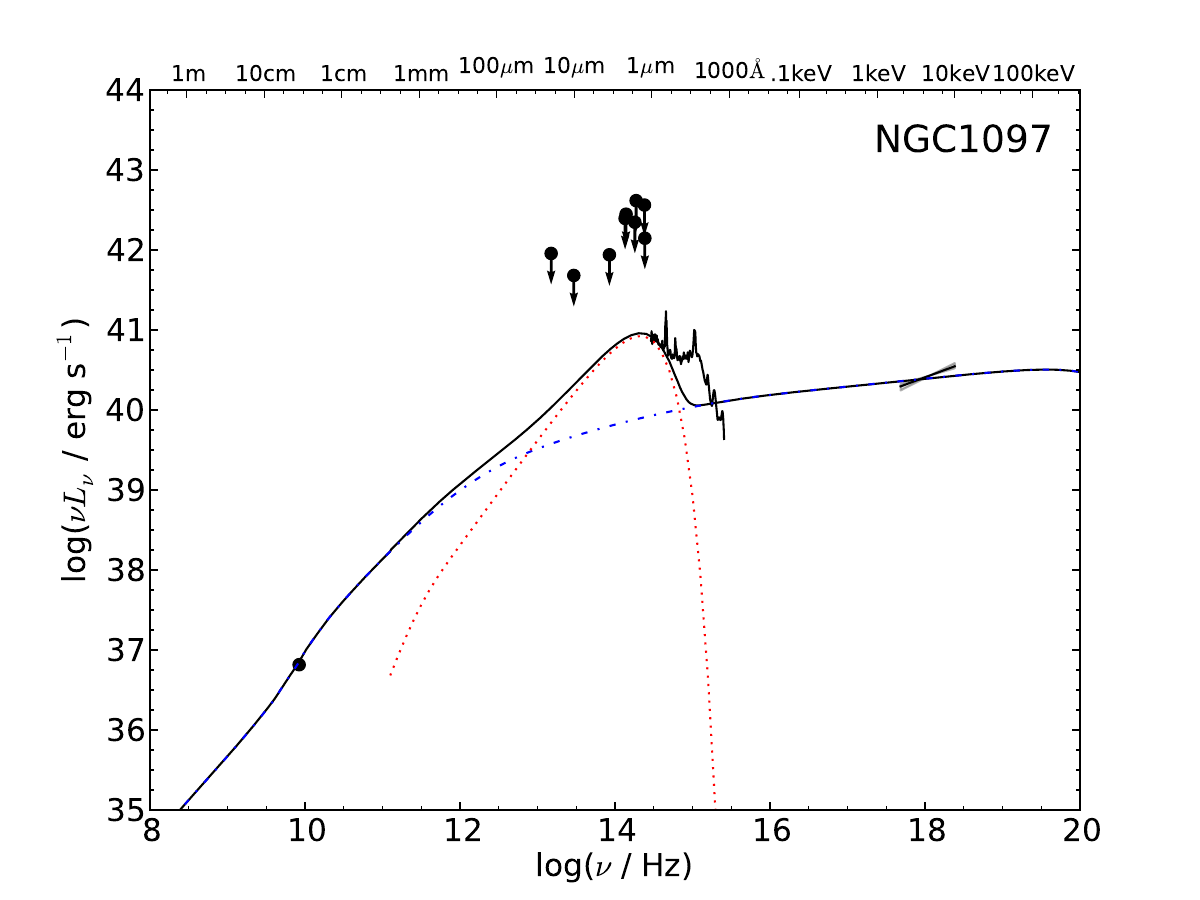}
}
\caption{Models for the SED of NGC 1097. The dashed, dotted and dot-dashed lines show respectively the emission from the ADAF, truncated thin disk and jet, while the solid line shows the sum of all the components. \textbf{Left:} model in which the ADAF dominates the observed X-ray emission.  \textbf{Right:} model in which the jet dominates the X-ray output, in which the ADAF contribution is negligible. For the model parameters, see \cite{nemmen12}.}
\label{1097}
\end{figure}

One nice example of LLAGN SED -- besides that of our Galactic Center, Sgr A* \cite{yuan03,yuan07}, which I will not discuss here -- is NGC 1097. In order to put the ADAF + truncated disk + jet scenario to the test, \cite{nemmen06} modeled the SED of NGC 1097 and obtained the fit displayed in Fig. \ref{1097}a (see also \cite{nemmen12}), where the contribution of the ADAF, truncated disk and jet were disentangled. This work demonstrates how the SEDs of LLAGNs can be important tools to constrain the physical properties of the central engines of nearby AGNs such as the mass accretion $\dot{M}$ and transition radius between the ADAF and thin disk. 

The sample of SEDs compiled by \cite{erac10} provides a great opportunity to extend the above kind of SED modeling to the LINER population as a whole. \cite{nemmen12} modeled the SEDs of a sample of 21 LINERs which were selected from the larger sample of \cite{erac10} following the criteria of availability of black hole mass estimates and good X-ray spectra. This modeling work -- with a careful exploration of the parameter space of the ADAF, jet and truncated disk models -- highlighted some of the limitations of the models as well as the limitations of the available SEDs as a tool to constrain the properties of the central engines of LLAGNs. We present some of the limitations in the next section.

\section{Some unresolved issues}

\begin{figure}
\center
\includegraphics[scale=0.55]{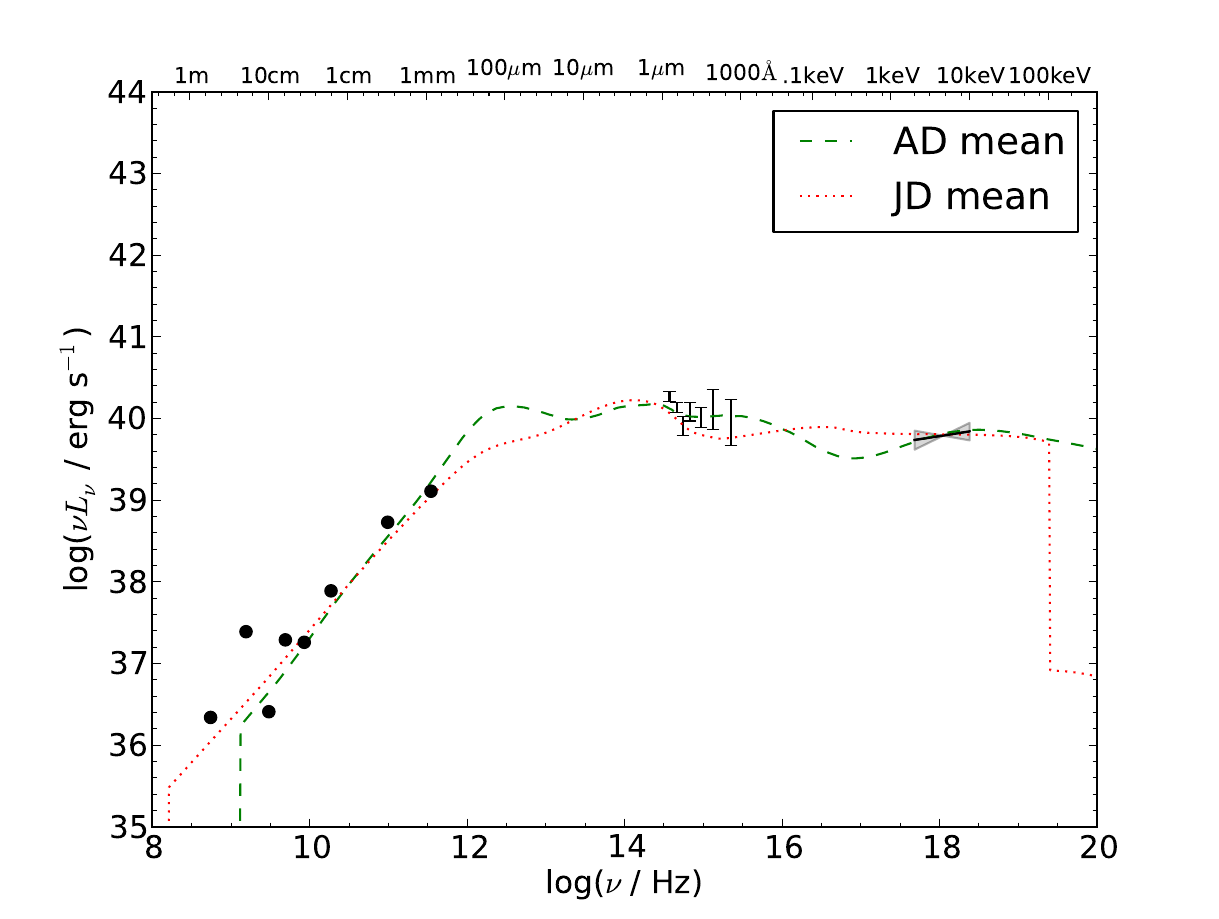}
\caption{The data points correspond to the median SED of LINERs obtained by \cite{erac10}. The dashed and dotted lines are two different averages of the models computed by \cite{nemmen12}. }
\label{avgsed} 
\end{figure}

The holes in our knowledge about the physics of the central engines of LLAGNs is directly connected to how well (or how badly) their multiwavelength SEDs are sampled, which is illustrated in Figure \ref{avgsed} (see also Fig. 4 in \cite{younes12}). We can see that, on average, LLAGN SEDs are poorly constrained, especially in mm, infrared (IR) and at energies $>10$ keV (compare, for instance, with the situation for quasars -- Fig. 7 in \cite{shang11} -- or blazars -- see Fig. 2 in \cite{ghise11}). \textbf{This poor multiwavelength sampling of LLAGNs constitutes, by itself, a compelling case for more observations and a better coverage of the SED.}

The need for a better sampling of the SEDs of LLAGNs is imperative in order to make progress towards answering the following important questions about the physics of LLAGNs:
\begin{itemize}
\item Can we find the signature of a truncated thin accretion disk in the IR?
\item Does the dusty toroidal structure disappear in LLAGNs?
\item Are the X-rays and higher energy radiation coming mostly from the ADAF or jet?
\item Can we find an unequivocal smoking gun for the ADAF emission?
\end{itemize}
For the rest of the review, I will focus on the first three points above.

\subsection{Testing accretion disk and dusty torus models with mid-infrared observations}

One unresolved issue in the study of LLAGNs is the presence of a truncated thin accretion disk. There is good observational support for the presence of a truncated disk in black hole binaries in the hard state \cite{yuan05,nar08}.
In LLAGNs, truncated disks with truncation radii $R \gtrsim 100 R_S$ and $\dot{M} \gtrsim 0.01 \dot{M}_{\rm Edd}$ would have emission peaking typically in the mid-IR \cite{nemmen06,yu11,nemmen12}. Until recently the available large-aperture IR data has been not sufficient to constrain the presence or properties of a truncated disk in LLAGNs \cite{yu11,nemmen12}.

Secondly, there is the unresolved issue of how LLAGNs fit into the unified framework of AGNs which invokes a toroidal dusty structure in order to explain the difference between types 1 and 2. It is not clear whether the torus disappears in the low mass accretion rate regime (e.g. \cite{mason12}). Interestingly enough, the torus signatures -- thermal emission and silicate emission and absorption features -- are precisely in the IR \cite{anton93}.

The search for the signatures of truncated thin accretion disks and the vanishing dusty torus prompts a better exploration of the IR band in LLAGNs. Fortunately, progress in this respect is being made (see the contributions of Perlman and Asmus in these proceedings; also  \cite{asmus11,mason12}). In particular, the comparison of how well the clumpy torus \cite{asensio09} and truncated accretion disk models explain the new, high-resolution, mid-IR imaging observations of \cite{mason12} will provide valuable insights on the issues above (Mason et al., in preparation).

\subsection{Where is the high-energy emission coming from?}

Both the ADAF and the jet are plausible sources of high-energy photons in LLAGNs, especially X-rays (e.g., \cite{merloni03,markoff08,yuan09,yu11,plotkin12,nemmen12}). Figure \ref{1097}a shows a SED fit for NGC 1097 in which the ADAF is dominant source of X-rays. Correspondingly, Figure \ref{1097}b shows a plausible fit in which the base of the jet dominates the high-energy emission \cite{nemmen12}. Which channel nature actually chooses to output X-rays in underfed supermassive black holes: ADAF or jet? 

ADAF and jet models predict different characteristic radio and X-ray variability timescales (e.g., \cite{ptak98}). Thus, one promising direction in the future to test which component  of the flow is dominant in X-rays is to carry out simultaneous monitoring of the variability of radio and X-rays \cite{miller10}. 

If the jet is energetically important in LLAGNs, it should be a source of $\gamma$-rays \cite{takami11} similarly to the case of M87 \cite{m87fermi}. The ADAF is also a potential $\gamma$-radiator \cite{maha97,oka03} though it is likely to contribute less than the jet in $\gamma$-rays (Nemmen et al., in preparation). Thanks to \emph{Fermi}-LAT, $\gamma$-rays are an exciting new window to study the relative contribution of the ADAF vs jet at high energies in LLAGNs. 

Finally, observations in the mm and sub-mm are also quite helpful because they constrain the ADAF synchrotron emission which is correlated with the X-ray emission. This follows because the synchrotron photons in the ADAF are inverse-Compton-scattered to X-rays. In this respect, ALMA could be a game changer.

\section{Conclusions}

LLAGNs are very interesting beasts which provide laboratories to study low-state black hole accretion and its consequences in nearby galaxies.  The SEDs of LLAGNs have been, until recently, poorly sampled in the cm and IR wavebands. Now the observational picture is changing, with new high-resolution observations in the IR being obtained for several galaxies. These observations will help to constrain the properties of truncated thin accretion disk and to test whether an obscuring dusty toroidal structure is present at all in such systems. Variability observing campaigns as well as $\gamma$-ray observations can help us understand the dominant component -- jet or ADAF -- responsible for the bulk of the high-energy radiation.

%
%
\small  
%
\section*{Acknowledgments}   
%
My research was supported by an appointment to the NASA Postdoctoral Program at Goddard Space Flight Center, administered by Oak Ridge Associated Universities through a contract with NASA. 
%

%
\end{document}